\documentstyle[12pt,aps,amsfonts,epsf]{revtex}

\def\upcite#1{\mbox{$\!^{\cite{#1}}$}}

\newcommand{\bra}[1]{\mbox{$\langle{#1}|$}}
\newcommand{\ket}[1]{\mbox{$|{#1}\rangle$}}

\def\I{{\rm i}}

\def\e{{\rm e}}

\begin{document}
\draft
\title{
Quantum CPU and Quantum Algorithm
\thanks{Supported by the National Natural Science Foundation of China under Grant No. 69773052 and the Fellowship of China Academy of Sciences}}

\author{An Min WANG$^{1,2,3}$}
\address{CCAST(World Laboratory) P.O.Box 8730, Beijing 100080$^1$\\
and Laboratory of Quantum Communication and Quantum Computing\\
University of Science and Technology of China$^2$\\
Department of Modern Physics, University of Science and Technology of China\\
P.O. Box 4, Hefei 230027, People's Republic of China$^3$}


\maketitle

\vspace{0.2in}

\begin{abstract}
{\em Making use of an universal quantum network -- QCPU proposed by me\upcite{My1}, it is obtained that the whole quantum network which can implement some the known quantum algorithms including Deutsch algorithm, quantum Fourier transformation, Shor's algorithm and Grover's algorithm}. 
\end{abstract}
\medskip



Quantum computing\upcite{Feynman,Deutsch} has attracted a lot of physicists since Shor found his quantum algorithm of factorization of a large number.\upcite{Shor} This is because that Shor's algorithm can speed up exponentially the factorization computation in a quantum computer than doing this in a classical computer. From then, quantum computer appears very promising and quantum algorithm plays a vital role. Actually, the high efficiency of quantum computer is just profited by the good quantum algorithms. 

Both classical and quantum algorithms consist of a series of quantum computing steps. In the quantum algorithm, a computing step can be a transformation or a quantum measurement. The final quantum computing task made of these steps is represented usually by the summation and product of them, which formed an total unitary transformation. In order to implement a given quantum algorithm including quantum simulating procedure, one needs to design, assemble and scale these quantum computing steps so that one finally can obtain a whole quantum network. 

Roughly speaking, the quantum network can be divided into three levels. First, the fundamental quantum network consists of a few basic quantum gates. Usually it can be constructed in the different methods. Secondly, the simple quantum network corresponds to a given computing step, which consists of the quantum gates and/or the fundamental quantum networks acting synchronously in time. The last is the complicated quantum network is made of the fundamental and simple quantum networks. 
A quantum algorithm is usually implemented by a complicated quantum network. 

As is well known, Barenco {\it et.al}'s proposed a beautiful method to design a quantum network in terms of the elementary gates.\upcite{Barenco} However, so far one has not found how to assemble the quantum network for the summation of simultaneous  transformations into a multiplication form. For example, although we have the quantum network for quantum Fourier transformation,\upcite{Ekert} it only can act on one state. Moreover, for the product of a series of successive transformations such as the time evolution operator, one doesn't know how to scale them into a whole quantum network. 
At present, except for the simplest Deustch's algorithm, I have not seen there is the whole quantum networks for the known main quantum algorithms. It is perhaps difficult to arrive at this aim only by Barenco {\it et.al}'s method. Thus, it seems to me that it is worth finding the new method since the integrality of the quantum network is necessary in practice. In this letter, by making use of an universal quantum network -- QCPU proposed by me,\upcite{My1} I obtain the whole quantum network which can implement some the known quantum algorithms including Deutsch algorithm,\upcite{Deutsch} quantum Fourier transformation, Shor's algorithm\upcite{Shor} and Grover's algorithm.\upcite{Grover}  In fact, QCPU also can be used in quantum simulating.\upcite{My2}

An universal quantum network for quantum computing $U$ is defined as:\upcite{My1}
\begin{equation}
Q(U)=\prod_{m,n=0}^{2^k-1}\exp\{(U_{mn}\ket{m}\bra{n}\otimes I_A)\cdot C_A^\dagger\}\label{UQN}
\end{equation}
where $I_R$ and $I_A$ are the identity matrices in the register space and the auxiliary qubit space respectively, and $C_A^\dagger=I_R\otimes c_A^\dagger=I_R\otimes \ket{1}{}_A{}_A\bra{0}$. 
If the graphics rules for the factor with form $\exp\{(U_{mn}\ket{m}\bra{n})\cdot C_A^\dagger\}$ are given out, the picture of quantum network $Q(U)$ can be drawn easily. It is easy to verify that $Q(U)$ can implement a general quantum computing. In this sense, it can be called the Quantum CPU. It is the most important that $Q(U)$ has two very useful new properties
\begin{equation}
Q(U_1+U_2+\cdots+U_r)=Q(U_1)Q(U_2)\cdots Q(U_r),
\end{equation}
\begin{equation}
Q(U_1U_2\cdots U_r)= I_R\otimes I_A+ C_A^\dagger\left(\prod_{j=1}^{r} C_AQ(U_j)\right) C_AC_A^\dagger,\label{CQC}
\end{equation} 
where $C_A$ is the so-called ``{\it Connector}" defined by  $
C_A=I_R\otimes c_A=I_R\otimes \ket{0}{}_A{}_A\bra{1}$. It is used to the preparing transformed state so that this prepared state can be used in the successive transformation. 
Furthermore, note that there are the relations $c_A^2=c_A^{\dagger 2}=0; c_A c_A^\dagger+c_A^\dagger c_A=I_A$, 
thus $c_A$ and $c_A^\dagger$ can be thought of as the fermionic annihilate and create operators respectively in the auxiliary qubit. In order to give out the realization of QCPU for the product of transformation in a form of full multiplication, eq.(\ref{CQC}) can be rewritten as
\begin{equation}
\bar{Q}(U_1U_2\cdots U_r)=(I_R)_{input}\otimes \left[C_A^\dagger\left(\prod_{j=1}^{r} C_AQ(U_j)\right) C_AC_A^\dagger\right]_{out},\label{CQCP}
\end{equation}
while the initial state is now prepared as $(\ket{\Psi(t)})_{input}\otimes (\ket{\Psi(t)}\otimes \ket{0}_A)_{out}$.
Therefore, it seems to me this new construction of the universal quantum network is scalable easily. 
 
This QCPU is universal because it can implement a general quantum computing task, including the general quantum algorithm and the quantum simulating procedure. This QCPU and its realizations are standard and easy-assemble because they only have two kinds of basic elements and two auxiliary elements. This QCPU and its realizations are scalable because they can easily connect each other. This QCPU is favor to the design of quantum algorithm including quantum simulating because it gives out the standard and explicit realization for each computing step. This QCPU is possible helpful for programming it since its simplicity in design. In terms of the QCPU, it is easy to obtain the whole quantum networks for the known main quantum algorithms as seen in the following.


Deutsch's problem is the first quantum algorithm and is the simplest one. However, it explains some important properties in quantum computing. Only taking one input qubit $x$ and one output qubit $f(x)$ (unknown function), we altogether have four possible values of the function $ f_1(0)=f_1(1)=0; f_2(0)=f_2(1)=1; f_3(0)=0$ and $f_3(1)=1;f_4(0)=1$ and $ f_4(1)=0$, because each $f(0)$ and $f(1)$ have two possible values. This problem is to judge whether $f$ is constant ($f(0)=f(1)$, such as $f_1$ or $f_2$) or balanced ($f(0)\neq f(1)$, such as $f_3$ or $f_4$). Intuitively, the best classical strategy is to calculate $f$ explicitly for input $0$ and $1$, and then to compare them. Therefore, in order to solve this problem, one has to two times calculations for the function $f$. However, quantum algorithm only needs single calculation to get the result. This algorithm was first proposed by Deutsch.\upcite{Deutsch} Now it has been improved and extended.\upcite{Ekert} Deutsch's algorithm can be described by the following four steps. First,   
prepare the first and second qubits respectively in $\ket{0}$ and $\ket{1}$. The total quantum state is $\ket{01}$. Secondly,  
act on each qubit by Hadamard transformation $H=(\sigma_1+\sigma_3)/\sqrt{2}$, where $\sigma_1,\sigma_3$ are usual Pauli matrix. Thus, it leads that $H\otimes H\ket{01}=(\ket{00}-\ket{01}+\ket{10}-\ket{11})/2$.
Thirdly, define a two-qubit gate $U_f: 
\ket{i,j}\rightarrow \ket{i,j\oplus f(i)}$, where $i,j=0,1$ and $\oplus$ means addition and mod 2. It acts on the above superposition state.
The last, again act on each qubit by Hadamard transformation. It is easy to verify that the final state is $\ket{01}$ (if $f=f_1$); 
$-\ket{01}$ (if $f=f_2$); $\ket{11}$ (if $f=f_3$); $-\ket{11}$ (if $f=f_4$).
Therefore, to make a measurement to the first qubit will reveal if the function is constant (output 0) or balanced (output 1). 

Obviously, Deutsch's algorithm can be written as the following:
\begin{eqnarray}
U({\rm Deutsch})&=& \left(
\begin{array}{cccc}
1&0&0&0\\
0&(-)^{f(0)}\delta_{f(0)f(1)}&0&\epsilon_{f(0)f(1)}\\
0&0&1&0\\
0&\epsilon_{f(0)f(1)}&0&(-)^{f(0)}\delta_{f(0)f(1)}
\end{array}\right)\\[10pt]
&=&\left(\begin{array}{cc}
1&0\\
0&1\end{array}\right)\otimes 
\left(\begin{array}{cc}
1&0\\
0&(-)^{f(0)}\delta_{f(0)f(1)}\end{array}\right)+
\left(\begin{array}{cc}
0&1\\
1&0\end{array}\right)\otimes 
\left(\begin{array}{cc}
0&0\\
0&\epsilon_{f(0)f(1)}\end{array}\right), \label{QCPU1}
\end{eqnarray}
where $\delta_{f(0)f(1)}= 1$ if $f(0)=f(1)$, otherwise it is zero, and $\epsilon_{f(0)f(1)}$ is an usual antisymmetric tensor defined by $\epsilon_{01}=-\epsilon_{10}=1$, the others are zero. Furthermore, Deutsch's algorithm can be simplified as an unitary transformation acting on the single qubit:
\begin{equation}
V({\rm Deutsch})=\left(\begin{array}{cc}
(-)^{f(0)}\delta_{f(0)f(1)}&\epsilon_{f(0)f(1)}\\
\epsilon_{f(0)f(1)}&
(-)^{f(0)}\delta_{f(0)f(1)}\end{array}\right).
\end{equation}
Thus, from this matrix it follows that eq.(\ref{QCPU1}) is equivalent to the realization of QCPU
\begin{equation}
Q(V({\rm Deutsch}))=\prod_{m,n=0}^1 \exp\{[V_{mn}({\rm Deutsch})\ket{x_m}\bra{x_n}\otimes I_A]\cdot C^\dagger\}.
\end{equation}
Its action on $\ket{1}\otimes \ket{0}_A$ will give the expected result. To make a measurement in basis $\ket{11}$ will reveal if the function is constant (output 0) or balanced (output 1).     


Quantum Fourier transformation plays an important role in quantum algorithm including factorization, search and simulating.\upcite{Ekert} Its matrix $F$ reads
\begin{equation}
F=\frac{1}{\sqrt{N}}\sum_{m,n=0}^{N-1}\e^{2\pi\I mn/N}\ket{x_m}\bra{p_n},\quad F^{-1}=\frac{1}{\sqrt{N}}\sum_{m,n=0}^{N-1}\e^{-2\pi\I mn/N}\ket{p_m}\bra{x_n}.
\end{equation}
Quantum Fourier transformation can be rewritten as
\begin{equation}
F=\sum_{n=0}^{2^k-1}\prod^k_{j=1}\frac{1}{\sqrt{2}}(\ket{0}+\e^{2^j\I\pi n/(2^k-1)}\ket{1})\bra{p_n}=\sum_{n=0}^{2^k-1}\prod_{\otimes,j=1}^k B_j(2^j\pi  n/(2^k-1)H_j\ket{0}\bra{p_n}.
\end{equation}  
From the definition of QCPU (\ref{UQN}), it is easy to get
\begin{equation}
Q(F)=\prod_{n=0}^{2^k-1}Q[B(n)HM_{0n}]= \prod_{m=0}^{2^k-1}\prod_{n=0}^{2^k-1}\exp\{[(B(n)H)_{m0}\ket{m}\bra{n}\otimes I_A] \cdot C^\dagger\},
\end{equation}
where $B(n)H=\prod_{\otimes,j=0}^{2^k-1}B_j[2^j\pi n/(2^k-1)]]H_j$ and $M_{0n}=\ket{0}\bra{x_n}$. 


Shor's algorithm is used to the factorization of a large number $N$. It can speed up exponentially computing in a quantum computer than doing this in a classical computer. Shor's algorithm can be described by the following five steps. First, 
start with two $k-$qubit registers in $\ket{0}\ket{0}$, then prepare the first register into a superposition with the equal weight in terms of Fourier transformation or $k-$qubit Hadamard gate denoted by $H$:
\begin{equation}
H\ket{0}\ket{0}=\sum_{n=0}^{2^k-1}\ket{n}\ket{0}.
\end{equation}
Secondly, select randomly a factor $a$ and make a mapping
\begin{equation}
\sum_{n=0}^{2^k-1}\ket{n}\ket{0}\stackrel{G}{\longrightarrow} \sum_{n=0}^{2^k-1}\ket{n}\ket{a^n {\rm mod} N}.
\end{equation}
Obviously
\begin{equation}
G=\sum_{n=0}^{2^k-1}\ket{n}\ket{a^n {\rm mod}N}\bra{n}\bra{0}.
\end{equation}
Thirdly, measure the second register by $I_1\otimes \ket{a^m {\rm mod}N}\bra{a^m {\rm mod}N}$ and obtain the result: $\sum_{j=0}^{[2^k/r]-1} |jr + l\rangle |u\rangle$. 
Fourthly, do Fourier transformation $F$ to the first register so that
\begin{equation}
U_{\rm DFT}\ket{jr+l}= \frac{1}{\sqrt{2^{k}}}\sum_{y=0}^{2^k-1} \exp \{2\pi {\rm i}(jr+l)y/2^k\} \ket{y}.
\end{equation}
and obtain the final state $\displaystyle \frac{1}{\sqrt{r}}\sum_{m=0}^{r-1} \exp (2\pi {\rm i}lm/r) |m2^k/r\rangle $. The last, measure the first register in the basis $y=m2^k/r$. If one obtains one values $y$, then solve equation $y/2^k=m/r$ to find the period. Once $r$ is known the factors of $N$ are obtained by calculating the greatest common divisor of $N$ and $a^{r/2}\pm 1$. 

Thus, the main steps in Shor's algorithm can be represented by one total transformation matrix:
\begin{equation}
U({\rm Shor})= (F\otimes I_2)M(a^m {\rm mod}N)GH;\quad M(n)=I_1\otimes \ket{n}\bra{n}.
\end{equation}
The product of the serval matrices is an easy problem. After we know the form of $U({\rm Shor})$, we are able to obtain the parameters to determine the realization of QCPU which can implement Shor's algorithm. But, it is more convenient, sometime more efficient, to connect the several realizations of QCPU for quantum computing steps and quantum measurement corresponding this algorithm together. Since we have had the realization of QCPU for quantum Fourier transformation, we only need the realizations of QCPU for $H$ and $G$. Obviously, starting from the initial state $\ket{0}_1\otimes\ket{0}_2\ket{0}_A$, it follows that $Q(H)$ is
\begin{equation}       
Q(H)= C_A^\dagger\prod_{j=1}^k (CQ(H_j\otimes I_2))\cdot C_A C_A^\dagger.
\end{equation}
Note that $H_j$ is a Hadamard transformation only acting on the $j-$ qubit in the first register and the second register has kept the original state. We do not need to add the other input register. The realization of QCPU for Mapping $G$ can read
\begin{equation}
Q(G)=\prod_{n=0}^{2^k-1}\exp\{(\ket{n}\ket{a^n {\rm mod}N}\bra{n}\bra{0})\otimes I_A)\cdot C^\dagger\}.
\end{equation}
Therefore the whole quantum network for Shor's factorization can be obtained:
\begin{equation}
\bar{Q}({\rm Shor})=(I_R)_{input}\otimes\left(C_A^\dagger C_AQ(F\otimes I_2)M(a^m{\rm mod}N)\otimes I_AC_A Q(G)C_A Q(H) C_AC_A^\dagger\right)_{out}.
\end{equation} 
From this example about Shor's algorithm, it seem to me to design a quantum algorithm is to seek an appropriate summation and product form of a series of simultaneous and successive quantum transformations as well as quantum measurements, and to implement a quantum algorithm is to give the realizations of QCPU for every transformations and then connect them including quantum measurements together. It is so standard and convenient.


Grover's algorithm is used to search the expected term in the unstructured data. It can be described by the following four steps. First, start with a $k-$qubit registers in $\ket{0}$, then prepare it into a superposition with the equal weight in terms of Fourier transformation or $k-$Hadamard gate, that is $H\ket{0}=\sum_{n=0}^{2^k-1}\ket{n}$. 
Secondly, do a reflection:
\begin{equation}
R_2=I-2\ket{j}\bra{j}=\sum_{n=0}^{2^k-1}(-1)^{\delta_{jm}}\ket{x_m}\bra{x_m},
\end{equation}
where $j$ corresponds to the expected data. Thirdly, 
make the following operation:
\begin{equation}
R_1=F^{-1}R_0F=F^{-1}[2\ket{0}\bra{0}-I]F=-F^{-1}\sum_{m=0}^{2^k-1}(-)^{\delta_{0 m}}\ket{x_m}\bra{x_m}F,
\end{equation} 
where $I$ is an identity matrix, $F$ is a quantum Fourier transformation, $F^{-1}$ is its inverse and $R_0=2\ket{0}\bra{0}-I$. The last, repeat $R_1R_2$ $\sqrt{N}\pi/4$ times and then do all measurements. 

Since the quantum network for quantum Fourier transformation has been obtained and its inverse has the similar realization but its parameters with a negative sign. While $R_0$ and $R_2$ is diagonal, it is very easy to get from our definition of QCPU
\begin{eqnarray}
Q(R_0)&=&\exp\{(2\ket{0}\bra{0}\otimes I_A)\cdot C^\dagger\}\exp\{-C_A^\dagger\}\\ 
Q(R_2)&=&Q(I_R)\exp\{(-2\ket{x_j}\bra{x_j}\otimes I_A)\cdot C^\dagger\}\exp\{C_A^\dagger\}.
\end{eqnarray}
Thus the quantum network for Grover's algorithm is just obtained
\begin{equation}
Q({\rm Grover})=(I_R)_{input}\otimes\left(C_A^\dagger C_AQ(F^{-1})CQ(R_0)CQ(F)CQ(R_2)CQ(H)C_AC_A^\dagger\right)_{out}.
\end{equation}
 
In conclusion, it is showed that the whole quantum networks of the known main quantum algorithms can be described by the realizations of QCPU. Actually, it is also able to obtain the quantum network to simulate Schr\"odinger equation. Therefore, QCPU is possible to play a useful role in searching for new quantum algorithm. This research is on progressing.
 
\bigskip
  
I would like to thank Artur Ekert for his great help and for his hosting my visit to center of quantum computing in Oxford University.

\end{document}